\documentclass{elsart}

\usepackage{epsf}

\usepackage{amssymb}

\usepackage{bm}

\begin{document}

\begin{frontmatter}



\title{van der Waals energy under strong atom-field coupling in doped carbon nanotubes}

\author{I.V. Bondarev\corauthref{cor1}} and
\ead{igor.bondarev@fundp.ac.be} \corauth[cor1]{Corresponding
author\\ On leave from the Institute for Nuclear Problems, The
Belarusian State University, Bobruiskaya Str. 11, 220050 Minsk,
BELARUS}
\author{Ph. Lambin}

\address{Facult\'{e}s Universitaires Notre-Dame de la Paix, 61
rue de Bruxelles, 5000 Namur, BELGIUM}

\begin{abstract}
Using a unified macroscopic QED formalism, we derive an integral
equation for the van der Waals energy of a two-level atomic system
near a carbon nanotube. The equation is valid for both strong and
weak atom-vacuum-field coupling. By solving it numerically, we
demonstrate the inapplicability of weak-coupling-based van der
Waals interaction models in a close vicinity of the nanotube
surface.
\end{abstract}


\end{frontmatter}

\section{Introduction}
\label{intro}

Carbon nanotubes (CNs) are graphene sheets rolled-up into
cylinders of approximately one nanometer in diameter.~Extensive
work carried out worldwide in recent years has revealed the
intriguing physical properties of these novel molecular scale
wires~\cite{Dai}.~Nanotubes have been shown to be useful for
miniaturized electronic, mechanical, electromechanical, chemical
and scanning probe devices and materials for macroscopic
composites~\cite{Baughman}.~Important is that their intrinsic
properties may be substantially modified in a controllable way by
doping with extrinsic impurity atoms, molecules and
compounds~\cite{Duclaux}.~Recent successful experiments on
encapsulation of single atoms into single-wall CNs~\cite{Jeong}
and their intercalation into single-wall CN
bundles~\cite{Duclaux,Shimoda} as well as the theory of
spontaneous decay in such systems~\cite{Bondarev02,Bondarev04}
stimulate an in-depth analysis of atom-CN van der Waals (vdW)
interactions.

The relative density of photonic states (DOS) near a CN
effectively increases due to the presence of additional surface
photonic states coupled with CN electronic quasiparticle
excitations~\cite{Bondarev02}. This causes an atom-vacuum-field
coupling constant (which is proportional to the photonic DOS) to
be very sensitive to an atom-CN-surface distance. In
particular~\cite{Bondarev04}, when the atom is close enough to the
CN surface and the atomic transition frequency is in the vicinity
of the resonance of the photonic DOS, the system shows
\emph{strong} atom-vacuum-field coupling giving rise to
rearrangement ("dressing") of atomic levels by vacuum-field
interaction. If the atom moves away from the CN surface, the
atom-field coupling strength decreases, smoothly approaching the
weak coupling regime at large atom-surface distances since the
role of surface photonic states diminishes causing the relative
density of photonic states to decrease.~This suggests strictly
\emph{nonlinear} atom-field coupling and a \emph{primary} role of
the distance-dependent (local) photonic DOS in the vicinity of the
CN, so that vacuum-QED-based vdW interaction models as well as
those based upon the linear response theory (see~\cite{Buhmann}
for a review) are in general inapplicable for an atom in a close
vicinity of a carbon nanotube.

To give this issue a proper consideration, we have developed a
simple quantum mechanical approach to the vdW energy calculation
of a (two-level) atomic system in the vicinity of an infinitely
long single-wall CN. The approach is based upon the perturbation
theory for degenerate atomic levels (see, e.g.,~\cite{Davydov}),
thereby allowing one to account for both strong and weak
atom-vacuum-field coupling regimes. In describing the atom-field
interaction, we follow a unified macroscopic QED formalism
developed for dispersing and absorbing media in Ref.~\cite{Welsch}
and adapted for an atom near a CN in
Refs.~\cite{Bondarev02,Bondarev04}. The formalism generalizes the
standard macroscopic QED normal-mode technique by representing
mode expansion in terms of a Green tensor of the (operator)
Maxwell equations in which material dispersion and absorbtion are
automatically included. In more detail, the Fourier-images of
electric and magnetic fields are considered as quantum mechanical
observables of corresponding electric and magnetic field
operators.~The latter ones satisfy the Fourier-domain operator
Maxwell equations modified by the presence of a~so-called operator
noise current written in terms of a bosonic field operator and an
imaginary dielectric permittivity.~This operator is responsible
for correct commutation relations of the electric and magnetic
field operators in the presence of medium-induced absorbtion.~The
electric and magnetic field operators are then expressed in terms
of a continuum set of operator bosonic fields by means of the
convolution of the operator noise current with the electromagnetic
field Green tensor of the system.~The bosonic field operators
create and annihilate single-quantum electromagnetic medium
excitations.~They are defined by their commutation relations and
play the role of the fundamental dynamical variables in terms of
which the Hamiltonian of the composed system "electromagnetic
field + dissipative medium" is written in a standard secondly
quantized form.

Using the approach summarized above, we derive an integral
equation for the vdW energy of a two-level atomic system near a
CN. The equation is represented in terms of the local photonic DOS
and is valid for both strong and weak atom-field coupling. By
solving it numerically, we demonstrate the inapplicability of
weak-coupling-based vdW interaction models in a close vicinity of
the CN.

\section{The van der Waals energy of an atom near a carbon
nanotube} \label{vdwenergy}

Consider a neutral atomic system with its centre-of-mass
positioned at the point $\mathbf{r}_{A}$ near an infinitely long
single-wall CN. Assign the orthonormal cylindric basis
$\{\mathbf{e}_{r},\mathbf{e}_{\varphi},\mathbf{e}_{z}\}$ with
$\textbf{e}_{z}$ directed along the CN axis. The total
nonrelativistic Hamiltonian of the whole system can then be
represented in the form (electric-dipole approximation, Coulomb
gauge, CGS units)~\cite{Bondarev04,Buhmann}
\begin{equation}
\hat{H}=\hat{H}_{F}+\hat{H}_{A}+\hat{H}^{(1)}_{AF}+\hat{H}^{(2)}_{AF}\,,
\label{Htot}
\end{equation}
where
\begin{equation}
\hat{H}_{F}=\int_{0}^{\infty}\!d\omega\hbar\omega\!\int\!d\mathbf{R}
\,\hat{f}^{\dag}(\mathbf{R},\omega)\hat{f}(\mathbf{R},\omega),
\label{Hf}
\end{equation}
\begin{equation}
\hat{H}_{A}=\sum_{i}{\hat{\mathbf{p}}^{2}\over{2m_{i}}}+
{1\over{2}}\sum_{i,j}{q_{i}q_{j}\over{|\mathbf{r}_{i}-\mathbf{r}_{j}|}},
\label{Ha}
\end{equation}
\begin{equation}
\hat{H}^{(1)}_{AF}=-\sum_{i}{q_{i}\over{m_{i}c}}\,\hat{\mathbf{p}}_{i}
\!\cdot\!\hat{\mathbf{A}}(\mathbf{r}_{A})+\hat{\mathbf{d}}\!\cdot\!
\nabla\hat{\varphi}(\mathbf{r}_{A}), \label{Haf1}
\end{equation}
\begin{equation}
\hat{H}^{(2)}_{AF}=\sum_{i}{q^{2}_{i}\over{2m_{i}c^{2}}}\,
\hat{\mathbf{A}}^{\!2}(\mathbf{r}_{A})
\label{Haf2}
\end{equation}
are, respectively, the Hamiltonian of the vacuum electromagnetic
field modified by the presence of the CN, the Hamiltonian of the
atomic subsystem, and the Hamiltonian of their interaction
(separated into two contributions according to their role in the
atom-CN vdW interaction -- see below). The operators
$\hat{f}^{\dag}$ and $\hat{f}$ in Eq.~(\ref{Hf}) are those
creating and annihilating single-quantum electromagnetic
excitations of bosonic type in the CN and the inner integral is
taken over the CN surface assigned by the vector
$\mathbf{R}=\{R_{cn},\phi,Z\}$ with $R_{cn}$ being the CN radius.
In Eqs.~(\ref{Ha})-(\ref{Haf2}), $m_{i}$, $q_{i}$,
$\hat{\mathbf{r}}_{i}$ and $\hat{\mathbf{p}}_{i}$ are,
respectively, the masses, charges, canonically conjugated
coordinates (relative to $\mathbf{r}_{A}$) and momenta of the
particles constituting the atomic subsystem,
$\hat{\mathbf{d}}=\sum_{i}q_{i}\hat{\mathbf{r}}_{i}$ is its
electric dipole moment operator. The vector potential
$\hat{\mathbf{A}}$ and the scalar potential $\hat{\varphi}$ of the
CN-modified electromagnetic field are given for an arbitrary
$\mathbf{r}=\{r,\varphi,z\}$ in the Schr\"{o}dinger picture by
\begin{equation}
\hat{\mathbf{A}}(\mathbf{r})=\int_{0}^{\infty}\!d\omega(i\omega)^{-1}
\underline{\hat{\mathbf{E}}}^{\perp}(\mathbf{r},\omega)+\mbox{h.c.},\;
-\nabla\hat{\varphi}(\mathbf{r})=\int_{0}^{\infty}\!d\omega\,
\underline{\hat{\mathbf{E}}}^{\parallel}(\mathbf{r},\omega)+\mbox{h.c.},
\label{vecscalpot}
\end{equation}
where
$\underline{\hat{\mathbf{E}}}^{\perp(\parallel)}(\mathbf{r},\omega)=
\int\!d\mathbf{r}^{\prime}\,\bm{\delta}^{\perp(\parallel)}
(\mathbf{r}-\mathbf{r}^{\prime})\cdot\underline{\hat{\mathbf{E}}}
(\mathbf{r}^{\prime},\omega)$ is the transverse (longitudinal)
electric field with $\delta^{\parallel}_{\alpha\beta}(\mathbf{r})=
-\nabla_{\alpha}\nabla_{\beta}(4\pi r)^{-1}$ and
$\delta^{\perp}_{\alpha\beta}(\mathbf{r})=\delta_{\alpha\beta}
\delta(\mathbf{r})-\delta^{\parallel}_{\alpha\beta}(\mathbf{r})$
being the longitudinal and transverse dyadic $\delta$-functions,
respectively, and $\underline{\hat{\mathbf{E}}}$ representing the
total electric field operator which satisfies the following set of
Fourier-domain Maxwell equations
\begin{equation}
\nabla\times\underline{\hat{\mathbf{E}}}(\mathbf{r},\omega)=
ik\,\underline{\hat{\mathbf{H}}}(\mathbf{r},\omega),
\label{Maxwell}
\end{equation}
\[
\nabla\times\underline{\hat{\mathbf{H}}}(\mathbf{r},\omega)=
-ik\,\underline{\hat{\mathbf{E}}}(\mathbf{r},\omega)+
{4\pi\over{c}}\underline{\hat{\mathbf{I}}}(\mathbf{r},\omega).
\]
Here, $\underline{\hat{\mathbf{H}}}$ stands for the magnetic field
operator, $k=\omega/c$, and
\begin{equation}
\underline{\hat{\mathbf{I}}}(\mathbf{r},\omega)=\!\int\!\!d\mathbf{R}\,
\delta(\mathbf{r}-\mathbf{R})\,\underline{\hat{\mathbf{J}}\!}\,(\mathbf{R},\omega)
=2\underline{\hat{\mathbf{J}}\!}\,(R_{cn},\varphi,z,\omega)\,\delta(r-R_{cn}),
\label{Irw}
\end{equation}
where
\begin{equation}
\underline{\hat{\mathbf{J}}\!}\,(\mathbf{R},\omega)=
\!\sqrt{\hbar\omega\mbox{Re}\sigma_{zz}(\mathbf{R},\omega)\over{\pi}}\,
\hat{f}(\mathbf{R},\omega)\,\textbf{e}_{z}
\label{current}
\end{equation}
is the operator noise current density associated with CN-induced
absorbtion~\cite{Bondarev04}, $\sigma_{zz}$ is the CN surface
axial conductivity per unit length (in describing CN electronic
properties, we use the surface axial conductivity model,
neglecting the CN azimuthal current and radial
polarizability~\cite{Slepyan}).

From Eqs.~(\ref{Maxwell})-(\ref{current}) it follows that
\begin{equation}
\underline{\hat{\mathbf{E}}}(\mathbf{r},\omega)=
i{4\pi\over{c}}\,k\!\int\!\!d\mathbf{R}\,\mathbf{G}(\mathbf{r},\mathbf{R},\omega)
\!\cdot\!\underline{\hat{\mathbf{J}}\!}\,(\mathbf{R},\omega)
\label{Erw}
\end{equation}
(and $\underline{\hat{\mathbf{H}}}=(ik)^{-1}
\nabla\times\underline{\hat{\mathbf{E}}}$ accordingly), where
$\mathbf{G}$ is the Green tensor of the classical electromagnetic
field in the vicinity of the CN. The set of
Eqs.~(\ref{Htot})-(\ref{Erw}) forms a closed electromagnetic field
quantization formalism in the presence of dispersing and absorbing
media which meets all the basic requirements of a~standard quantum
electrodynamics~\cite{Welsch}. All information about medium (the
CN in our case) is contained in the Green tensor $\mathbf{G}$
whose components satisfy the equation
\begin{equation}
\sum_{\alpha=r,\varphi,z}\!\!\!
\left(\mathbf{\nabla}\!\times\mathbf{\nabla}\!\times-\,k^{2}\right)_{\!z\alpha}
G_{\alpha z}(\mathbf{r},\mathbf{R},\omega)=
\delta(\mathbf{r}-\mathbf{R}), \label{GreenequCN}
\end{equation}
together with the radiation conditions at infinity and the
boundary conditions on the CN surface. This tensor was derived and
analysed in Ref.~\cite{Bondarev04}.

Starting from Eqs.~(\ref{Htot})-(\ref{Haf2}) and using
Eqs.~(\ref{vecscalpot}),~(\ref{current}),~(\ref{Erw}), one obtains
the following secondly quantized Hamiltonian for a
\emph{two-level} atomic system interacting with the CN-modified
electromagnetic field
\begin{equation}
\hat{H}=\int_{0}^{\infty}\!d\omega\hbar\omega\!\int\!d\mathbf{R}
\,\hat{f}^{\dag}(\mathbf{R},\omega)\hat{f}(\mathbf{R},\omega)+
{\hbar\tilde{\omega}_{A}\over{2}}\,\hat{\sigma}_{z}
\label{Htwolev}
\end{equation}
\[
+\int_{0}^{\infty}\!d\omega\!\int\!d\mathbf{R}\left[
\mbox{g}^{(+)}(\mathbf{r}_{A},\mathbf{R},\omega)\,\hat{\sigma}^{\dag}-
\mbox{g}^{(-)}(\mathbf{r}_{A},\mathbf{R},\omega)\,\hat{\sigma}\right]
\hat{f}(\mathbf{R},\omega)+\mbox{h.c.}.
\]
Here, the Pauli operators $\hat{\sigma}_{z}\!=\!|u\rangle\langle
u|-|l\rangle\langle l|$, $\hat{\sigma}\!=\!|l\rangle\langle u|$,
$\hat{\sigma}^{\dag}\!=\!|u\rangle\langle l|$ describe electric
dipole transitions between the two atomic states, upper
$|u\rangle$ and lower $|l\rangle$, separated by the transition
frequency $\omega_{A}$. This (bare) frequency is modified by the
interaction (\ref{Haf2}) which, being independent of the atomic
dipole moment, does not contribute to mixing the $|u\rangle$ and
$|l\rangle$ states, giving rise, however, to the new
\emph{renormalized} transition frequency
\begin{equation}
\tilde{\omega}_{A}=\omega_{A}\left(1-{2\over{(\hbar\omega_{A})^{2}}}
\int_{0}^{\infty}\!d\omega\!\int\!d\mathbf{R}\,
|\mbox{g}^{\perp}(\mathbf{r}_{A},\mathbf{R},\omega)|^{2}\right)
\label{omegarenorm}
\end{equation}
in the second term of Eq.~(\ref{Htwolev}). On the contrary, the
interaction (\ref{Haf1}), being dipole moment dependent, mixes the
$|u\rangle$ and $|l\rangle$ states, yielding the third term of the
Hamiltonian (\ref{Htwolev}) with the interaction matrix elements
\begin{equation}
\mbox{g}^{(\pm)}(\mathbf{r}_{A},\mathbf{R},\omega)=
\mbox{g}^{\perp}(\mathbf{r}_{A},\mathbf{R},\omega)\pm
{\omega\over{\omega_{A}}}\,\mbox{g}^{\parallel}(\mathbf{r}_{A},\mathbf{R},\omega),
\label{gpm}
\end{equation}
where
\begin{equation}
\mbox{g}^{\perp(\parallel)}(\mathbf{r}_{A},\mathbf{R},\omega)=-i{4\omega_{A}\over{c^{2}}}
\sqrt{\pi\hbar\omega\mbox{Re}\sigma_{zz}(\omega)}\,d_{\alpha}G_{\alpha
z}^{\perp(\parallel)}(\mathbf{r}_{A},\mathbf{R},\omega)
\label{gperppar}
\end{equation}
with $G_{\alpha
z}^{\perp(\parallel)}(\mathbf{r}_{A},\mathbf{R},\omega)=
\int\!d\mathbf{r}\,\delta^{\perp(\parallel)}_{\alpha\beta}(\mathbf{r}_{A}-\mathbf{r})\,
G_{\beta z}^{\perp(\parallel)}(\mathbf{r},\mathbf{R},\omega)$. The
matrix element (\ref{gperppar}), being squared and integrated over
the nanotube surface, may be represented in the form
\begin{equation}
\int\!d\mathbf{R}\,|\mbox{g}^{\perp(\parallel)}(\mathbf{r}_{A},\mathbf{R},\omega)|^{2}=
\hbar^{2}{\Gamma_{0}(\omega)\over{2\pi}}\left(\omega_{A}\over{\omega}\right)^{\!2}
\xi^{\perp(\parallel)}(\mathbf{r}_{A},\omega),
\label{DOS}
\end{equation}
where $\xi^{\perp(\parallel)}$ is the transverse (longitudinal)
\emph{local} photonic DOS defined by~\cite{Bondarev04}
\begin{equation}
\xi^{\perp(\parallel)}(\mathbf{r}_{A},\omega)=
{\Gamma^{\perp(\parallel)}(\mathbf{r}_{A},\omega)\over{\Gamma_{0}(\omega)}}
\label{DOSdef}
\end{equation}
with $\Gamma^{\perp(\parallel)}(\mathbf{r}_{A},\omega)\!=\!8\pi
d_{\alpha}d_{\beta}\,\mbox{Im}^{\perp(\parallel)}
G_{\alpha\beta}^{\perp(\parallel)}(\mathbf{r}_{A},\mathbf{r}_{A},\omega)
/\hbar c^{2}$ being the transverse (longitudinal) atomic
spontaneous decay rate near the CN, and $\Gamma_{0}$ representing
the same quantity in vacuum where $\mbox{Im}G^{0}_{\alpha\beta}\!=
\delta_{\alpha\beta}\,\omega/6\pi c$. Note that
$\mbox{Im}^{\perp(\parallel)}G_{\alpha\beta}^{\perp(\parallel)}
\!=\mbox{Im}G^{0}_{\alpha\beta}+\mbox{Im}^{\perp(\parallel)}
\overline{G}_{\alpha\beta}^{\perp(\parallel)}$ with the second
term representing the "pure" CN contribution to the total
imaginary Green tensor, so that Eq.~(\ref{DOSdef}) may also be
written in the form
$\xi^{\perp(\parallel)}(\mathbf{r}_{A},\omega)=
1+\overline{\xi}^{\perp(\parallel)}(\mathbf{r}_{A},\omega)$.

\begin{figure}[t]
\hspace{-0.75cm}\epsfxsize=15.0cm\epsfbox{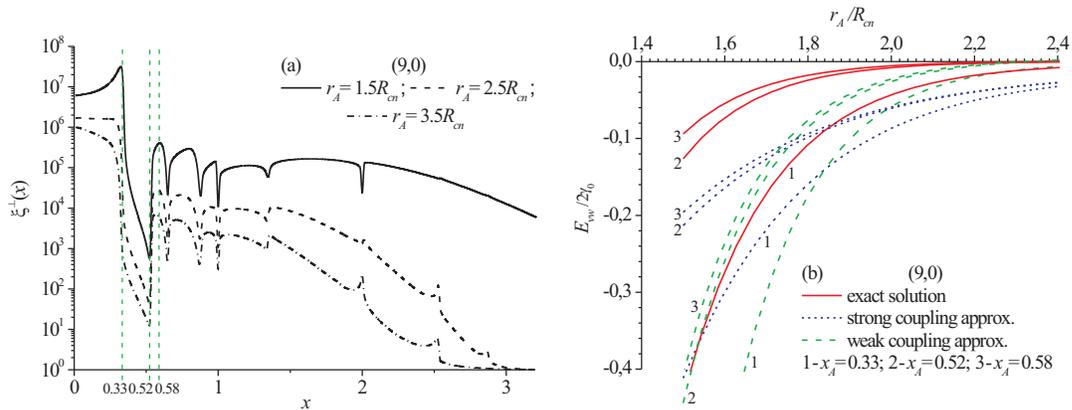}
\caption{Typical example of photonic DOS (the transverse local
photonic DOS) (a) and the vdW energy (b) of the atom outside the
(9,0) CN. The (bare) atomic transition frequencies are indicated
by dashed lines in Fig.~\ref{fig1}(a).} \label{fig1}
\end{figure}

The Hamiltonian (\ref{Htwolev}) is a starting point for the vdW
energy calculation. The latter is nothing but the
$\mathbf{r}_{A}$-dependent contribution to the ground-state energy
given by this Hamiltonian. Important is that the two-level atomic
subsystem (the second term in Eq.~(\ref{Htwolev})) is now
described by the \emph{renormalized} frequency (\ref{omegarenorm})
which is seen (see Eq.~(\ref{DOS})) to decrease with increasing
$\xi^{\perp}$ (i.e. when the atom approaches the CN
surface~\cite{Bondarev02,Bondarev04}), thereby bringing the two
atomic levels together, or even making them degenerated if
$\xi^{\perp}$ is large enough. To take this fact into account in a
correct way, one has to calculate the energy using the
perturbation theory for degenerated levels (see,
e.g.,~\cite{Davydov}). In so doing, one also has to account for
the upper state degeneracy of the whole system with respect to
$\mathbf{R}$ and $\omega$. In view of this, the wave function of
the whole system should be written in the form
\begin{equation}
|\psi\rangle=C_{l}\,|l\rangle|\{0\}\rangle+
\int_{0}^{\infty}\!d\omega\!\int\!d\mathbf{R}\,
C_{u}(\mathbf{R},\omega)\,|u\rangle|\{1(\mathbf{R},\omega)\}\rangle,
\label{wfunc}
\end{equation}
where $|\{0\}\rangle$ is the vacuum state of the field subsystem,
$|\{1(\mathbf{R},\omega)\}\rangle$ is its single-quantum excited
state, $C_{l}$ and $C_{u}$ are unknown mixing coefficients of the
lower and upper states of the \emph{whole} system. The total
ground-state energy $E$ is then given by the solution of a secular
equation obtained by applying the Hamiltonian (\ref{Htwolev}) to
the wave function (\ref{wfunc}). This yields the integral equation
\begin{equation}
E=-{\hbar\tilde{\omega}_{A}\over{2}}-
\int_{0}^{\infty}\!d\omega\int\!d\mathbf{R}\,
{\displaystyle|\mbox{g}^{(-)}(\mathbf{r}_{A},\mathbf{R},\omega)|^{2}
\over{\displaystyle\hbar\omega+{\hbar\tilde{\omega}_{A}\over{2}}-E}},
\label{E}
\end{equation}
which the vdW energy $E_{vw}$ is determined from by means of the
obvious relation $E=-\hbar\omega_{A}/2+E_{vw}(\mathbf{r}_{A})$
where the first term is the unperturbed ground-state energy of the
two-level atom. Using further the dimensionless variables
$\varepsilon_{vw}=E_{vw}/2\gamma_{0}$ and
$x={\hbar\omega/2\gamma_{0}}$ ($\gamma_{0}=2.7$~eV is the carbon
nearest neighbor hopping integral appearing in $\sigma_{zz}$ in
Eq.~(\ref{current})), one finally arrives, in view of
Eqs.~(\ref{omegarenorm})-(\ref{DOS}), at the following integral
equation
\begin{equation}
\varepsilon_{vw}(\mathbf{r}_{A})={\hbar\over{3\pi\gamma_{0}x_{A}}}
\int_{0}^{\infty}\!dx\Gamma_{0}(x)\overline{\xi}^{\perp}(\mathbf{r}_{A},x)
\label{Evw}
\end{equation}
\[
-{\hbar\over{3\pi\gamma_{0}}}\int_{0}^{\infty}\!dx
{\displaystyle\Gamma_{0}(x)\left[\overline{\xi}^{\perp}(\mathbf{r}_{A},x)+
{\left(x\over{x_{A}}\right)^{\!2}}\overline{\xi}^{\parallel}(\mathbf{r}_{A},x)\right]
\over{\displaystyle x+x_{A}-{\hbar\over{3\pi\gamma_{0}x_{A}}}
\int_{0}^{\infty}\!dx\Gamma_{0}(x)\overline{\xi}^{\perp}(\mathbf{r}_{A},x)
-\varepsilon_{vw}(\mathbf{r}_{A})}}
\]
describing the vdW energy in terms of the distance-dependent
photonic DOS analyzed in great detail in Ref.~\cite{Bondarev04}
and shown in Figure~\ref{fig1}(a) for a particular case of the
atom outside the (9,0) CN. Eq.~(\ref{Evw}) is \emph{universal} in
the sense that it covers both strong and weak atom-field coupling
regimes. These two limiting cases are obtained by putting
$x_{A}\approx(\hbar/3\pi\gamma_{0}x_{A})\int_{0}^{\infty}
\!dx\Gamma_{0}(x)\overline{\xi}^{\perp}(\mathbf{r}_{A},x)$ and
$\varepsilon_{vw}\approx0$, respectively, in the denominator of
the second term. The former is nothing but a degeneracy condition
of the unperturbed atomic levels due to the atom-field
interaction, taking place at small atom-CN-surface distances when
the local photonic DOS is large enough. The latter occurs at lower
photonic DOS values when the atom is not very close to the CN
surface. In this latter case, Eq.~(\ref{Evw}) gives rise to an
ordinary second order perturbation theory result with the total
Hamiltonian (\ref{Htwolev}) where the last term is considered as a
perturbation. This result is slightly "improved" compared with
those obtained from various weak-coupling-based models discussed
in the literature (see~\cite{Buhmann} and Refs. therein) since a
part of the atom-field interaction is included in the unperturbed
Hamiltonian via the renormalized transition frequency.

\section{Numerical results and discussion}
\label{numerics}

Using Eq.~(\ref{Evw}), we have simulated the vdW energy of the
atom outside the (9,0) CN. The local photonic DOS functions
$\overline{\xi}^{\perp,\parallel}(\mathbf{r}_{A},x)=
\xi^{\perp,\parallel}(\mathbf{r}_{A},x)-1$ were computed in the
same manner as it was done in Ref.~\cite{Bondarev04}. The
free-space spontaneous decay rate was approximated by the
expression $\Gamma_{0}(x)\approx\alpha^{3}2\gamma_{0}x/\hbar$
($\alpha=1/137$ is the fine-structure constant) valid for
hydrogen-like atoms~\cite{Davydov}. Figure~\ref{fig1}(a) shows the
transverse local photonic DOS, as a typical example, for the atom
located at several distances from the CN surface. The DOS function
is seen to increase with decreasing the atom-CN-surface distance,
representing the increase of the atom-field coupling strength as
the atom approaches the nanotube surface~\cite{Bondarev04}. The
vertical dashed lines indicate the bare atomic transitions
frequencies $x_{A}$ for which the vdW energies shown in
Figure~\ref{fig1}(b) were calculated. The frequencies are typical
for heavy hydrogen-like atoms such as Cs for which $x_{A}$ may be
estimated from its first ionization potential~\cite{Lide} to be
$3.89\mbox{\,eV}\times3/4\times(2\gamma_{0})^{-1}\sim0.5$ (the
factor $3/4$ comes from the Lyman series of~H), or less for highly
excited Rydberg states.

In Figure~\ref{fig1}(b), the vdW energies given by exact numerical
solutions of Eq.~(\ref{Evw}) are compared with those obtained from
the same equation within the weak and strong coupling
approximations. At small atom-CN-surface distances, the exact
solutions are seen to be fairly well reproduced by those obtained
in the strong coupling approximation, clearly indicating the
strong atom-vacuum-field coupling regime in a close vicinity of
the nanotube surface. The deviation from the strong coupling
approximation increases with $x_{A}$, that is easily explicable
since the degeneracy condition of the unperturbed atomic levels is
more difficult to reach for larger interlevel separations. As the
atom moves away from the CN surface, the exact solutions deviate
from the strong coupling solutions and approach those given by the
weak coupling approximation, indicating the reduction of the
atom-field coupling strength with raising the atom-surface
distance. The weak-coupling solutions are seen to be divergent
close to the nanotube surface as it should be because of the
degeneracy of the unperturbed (bare) atomic levels in this region.

To conclude, we have investigated vdW interactions of a two-level
atomic system near a single-wall CN within the unified QED
approach accounting for both strong and weak atom-vacuum-field
coupling. We have demonstrated the inapplicability of
weak-coupling-based vdW interaction models in a close vicinity of
the nanotube surface where the local photonic DOS effectively
increases giving rise to atom-field coupling enhancement followed
by the degeneracy and rearrangement ("dressing") of bare atomic
levels.

The work was performed within the framework of the Belgian
PAI-P5/01 project. I.B.~thanks the Belgian OSTC.




\end{document}